\documentclass[11pt]{article}

\usepackage[utf8]{inputenc}
\usepackage[T1]{fontenc}
\usepackage{amsmath,amssymb}
\usepackage{graphicx}
\usepackage{booktabs}
\usepackage[hypertexnames=false]{hyperref}
\usepackage{xurl}  
\usepackage{xspace}
\usepackage{microtype}
\usepackage{xcolor}
\usepackage[margin=1in]{geometry}

\hypersetup{
    colorlinks=true,
    linkcolor=blue,
    urlcolor=cyan,
    citecolor=blue,
    pdftitle={The Windows IOCTL Census: A Corpus-Scale, Multi-Architecture Database of the Driver Control-Code Surface},
    pdfauthor={Michael J.\ Bommarito II},
    pdfsubject={Vulnerability research, binary analysis, Windows drivers, IOCTL, LLM enrichment},
    pdfkeywords={Windows drivers, IOCTL, attack surface, binary analysis, taint analysis, LLM enrichment, BYOVD}
}

\newcommand{\nbins}{27{,}087\xspace}
\newcommand{\nx}{12{,}719\xspace}        
\newcommand{\nfns}{8{,}184{,}655\xspace}
\newcommand{\nfnsM}{8.18M\xspace}
\newcommand{\nedges}{15{,}950{,}899\xspace}
\newcommand{\nedgesM}{15.95M\xspace}

\newcommand{\ncodes}{849{,}048\xspace}

\newcommand{\nprimary}{10{,}781\xspace}   
\newcommand{\ndisp}{114{,}661\xspace}     
\newcommand{\nhandlers}{848{,}094\xspace}
\newcommand{\nsinks}{63{,}263\xspace}
\newcommand{\nenriched}{329\xspace}
\newcommand{\ncatalog}{26{,}111\xspace}
\newcommand{\xsixrecall}{83.4\%\xspace}         
\newcommand{\nxsixrec}{10{,}927\xspace}         
\newcommand{\nxsixcodes}{2.2M\xspace}           
\newcommand{\covbefore}{40\%\xspace}            
\newcommand{\covafter}{80\%\xspace}             
\newcommand{\nsurf}{21{,}708\xspace}            
\newcommand{\ncodesallK}{3.1M\xspace}
\newcommand{\nsamp}{1{,}000\xspace}
\newcommand{\nioccompleted}{802\xspace}
\newcommand{\nioccov}{126\xspace}
\newcommand{\niocvd}{54\xspace}
\newcommand{\ntaintd}{231\xspace}
\newcommand{\nboth}{35\xspace}
\newcommand{\ntonlynoh}{171\xspace}
\newcommand{\sys}{\textsc{Recover-Enrich-Rank}\xspace}
\newcommand{\glaurung}{\texttt{glaurung}\xspace}
\newcommand{\code}[1]{\texttt{#1}}

\title{\textbf{The Windows IOCTL Census}:\\
  A Corpus-Scale, Multi-Architecture Database\\
  of the Driver Control-Code Surface}

\author{
Michael J.\ Bommarito II\thanks{Portions of this work were prepared with
assistance from large language models. The author is solely responsible for
all content, including any errors or omissions. This work was conducted for
defensive and authorized vulnerability-research purposes; see the data-release
and ethics notes in Section~\ref{sec:discussion}.}\\
\texttt{michael.bommarito@gmail.com}
}

\date{June 2026}

\begin{document}
\maketitle

\begin{abstract}
A Windows driver exposes its kernel through I/O control (IOCTL) codes, and a
single unchecked length on the buffer behind one turns an unprivileged call into
a kernel write. The research community has strong \emph{scanners} for this
surface and a curated \emph{list} of known-bad drivers, but no map of the surface
itself. We build that map. The Windows IOCTL Census is a queryable database of the
control-code dispatch surface of \nbins\ signed Windows drivers, recovered by one
deterministic, architecture-neutral pass with no symbolic execution. Reading a
lifted intermediate representation instead of running a symbolic engine lets it
recover a dispatch surface for \covafter\ of the corpus across x86 and x64,
including the 32-bit half existing scanners abort on. On the 64-bit lane it adds
handler reachability, taint, and the call graph. An LLM ranks the reachable
handlers for triage. We release the census as a public dataset of tens of
millions of rows: \nbins\ binaries, \ncodesallK\ decoded control codes, \nfnsM\
functions, and \nedgesM\ call edges.
\end{abstract}

\section{Introduction}
\label{sec:intro}

The device-control path is one of the shortest routes from an unprivileged
Windows process to kernel memory. A user program calls
\code{DeviceIoControl} with a control code and an input buffer. The I/O manager
routes the request to the driver's \code{IRP\_MJ\_DEVICE\_CONTROL} handler, which
dispatches on the code and operates on the attacker-supplied buffer. If
the device's access-control list admits unprivileged callers and a handler
trusts a buffer length it does not check, the result is an out-of-bounds kernel
read or write. The same unchecked-handler pattern is what the ``bring your own
vulnerable driver'' (BYOVD) class~\cite{byovd} abuses from the other direction:
there an already-privileged attacker loads a signed but flawed driver to reach
such a handler, where here an unprivileged caller reaches one that ships on the
machine. The handler bug is shared. Only the route to it differs. One driver makes the
surface concrete: the display driver \code{nvlddmkm.sys}, whose device descriptor
grants write access to the World SID and exposes \code{METHOD\_BUFFERED} control
codes (device type \code{0x200}) to any caller. Its kernel-mode escape surface
has carried public privilege-escalation CVEs, among them CVE-2024-0090, an
out-of-bounds write on the display driver's kernel-mode escape surface. It is one of \nbins\
drivers in the census, which holds \ncodesallK\ decoded control codes,
\nhandlers\ handlers, and, on the 64-bit lane, \nfnsM\ functions and \nedgesM\
call edges.

Two bodies of work address this surface, and a gap sits between them. On one
side are \emph{scanners}: ScrewedDrivers~\cite{screweddrivers},
POPKORN~\cite{popkorn}, and the
angr-based \code{ioctlance}~\cite{ioctlance} symbolically execute a driver's
\code{DriverEntry}, locate the dispatch routine, and explore handler paths for
risky primitives. These tools run at corpus scale but emit a one-shot list of
findings and discard the structure they recovered along the way. They also
inherit symbolic execution's well-documented path explosion and false-negative
rate~\cite{popkorn}. Architecture is a further limit: in our runs \code{ioctlance} aborts on
a 32-bit driver, and roughly half of a shipping driver corpus is 32-bit x86, so
a large legacy tail, exactly where the BYOVD class often lives, is out of their
reach entirely. On the other side is \emph{curation}: LOLDrivers
\cite{loldrivers} catalogs a few hundred known-vulnerable and known-malicious
drivers with hashes, CVE mappings, and per-driver notes, and Microsoft ships a
vulnerable-driver blocklist. These are authoritative for what they cover, but
they are hand-built lists of the already-known, not a systematic extraction of
the IOCTL surface across the long tail of shipping drivers.

What is missing is the middle layer: a persistent, queryable record of the
\emph{surface} of an entire driver corpus, against which both kinds of work can
run. We build that layer, and we build it where the scanners cannot. Because our
recovery reads an architecture-neutral intermediate representation rather than
running a symbolic engine, one recognizer maps the control-code surface across
both x86 and x64, covering \covafter\ of the corpus including the 32-bit half
every symbolic scanner aborts on. Our contribution is a database, not a detector:
a single relational store, keyed by \code{(binary\_sha256, function\_va)}, that
holds for \nbins\ signed drivers the recovered dispatch routines and decoded
control codes, and on the deeply analyzed x64 lane the handler functions and
their reachability from the dispatcher, buffered-input taint sinks, the
inter-procedural call graph, imports and exports, and PE version and signing
provenance. On top of the
deterministic facts we attach a low-cost LLM verdict to the handlers
behind a permissive device descriptor, and we expose the whole as views that
rank the unaudited surface for a downstream human or agent.

Three commitments shape the design. First, \textbf{deterministic before
model}: dispatch recovery, code decoding, reachability, and taint are exact,
cacheable, and computed for every driver. The model sees a compact
feature summary of a reachable handler, never raw decompiled bytes, echoing how
recent target-selection work~\cite{needles,siftrank} selects what the model sees. Second, \textbf{the structure is the artifact}:
because the call graph, taint sinks, and decoded codes persist, the database
answers questions a finding list cannot, such as which drivers expose the same
control code, or which user-reachable handlers reach a copy sink without an
intervening length check. Third, \textbf{scoping}: the model's tiers are
prioritization hypotheses, and we report no true-positive rate. The
``critical'' rows are candidate leads for a verification pass.

This paper describes the method, shows that the deterministic recovery
generalizes across architectures to cover \covafter\ of the corpus where the
symbolic tools reach only the 64-bit half, characterizes where it stops (the
x86 lane is surface-only, and a tail of oversized binaries and other driver
models is recorded with provenance), and reports aggregate statistics. We release
the structural census as a public dataset and withhold the targeting tier;
Section~\ref{sec:discussion} states why.

\section{Method}
\label{sec:method}

\begin{figure}[t]
  \centering
  \includegraphics[width=\textwidth]{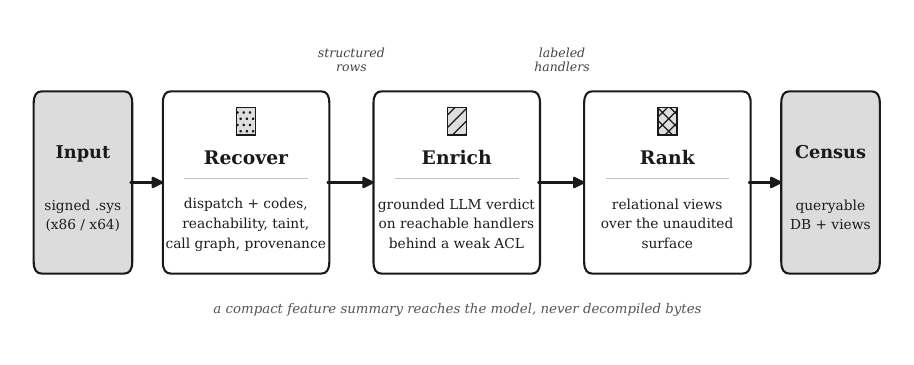}
  \caption{The \sys\ pipeline. \emph{Recover} extracts dispatch, decoded codes,
  handler reachability, controlled-input taint, the call graph, and provenance
  from a driver binary. \emph{Enrich} attaches a feature-grounded model verdict
  to the user-reachable handlers. \emph{Rank} serves views over the unaudited
  surface. Only a compact feature summary of a reachable handler reaches the
  model, never decompiled bytes.}
  \label{fig:pipeline}
\end{figure}

The pipeline runs as two deterministic-then-model passes over a driver corpus,
which we frame as three stages (Figure~\ref{fig:pipeline}) writing to a single
relational store keyed by \code{(binary\_sha256, function\_va)}. \emph{Recover}
extracts the IOCTL surface and program structure of each driver with no
symbolic execution. \emph{Enrich} attaches a feature-grounded model verdict to
the user-reachable handlers. \emph{Rank} exposes views that order the unaudited
surface. Table~\ref{tab:schema} gives the schema: the tables each stage writes,
the shared key, and the views. The corpus is \nbins\ signed drivers: \ncatalog\ harvested from the
Microsoft Update Catalog plus the inbox driver sets of three live Windows
builds.

\subsection{Recover}
\label{sec:recover}
The deterministic pass is in-process and per-driver, with no kernel, device, or
symbolic engine. For each binary it records, in order:

\paragraph{Dispatch and codes.} A dispatch-recovery backend identifies the
\code{IRP\_MJ\_DEVICE\_CONTROL} routine and the constants it compares the
control code against, recovering both \code{cmp}-immediate and jump-table
dispatch shapes. An index-computed fallback fires only when the primary returns
nothing. Each control code is decoded into its device type, transfer method
(\code{BUFFERED}/\allowbreak\code{IN\_DIRECT}/\allowbreak\code{OUT\_DIRECT}/\allowbreak\code{NEITHER}), required
access, and function number, and linked to its handler address.

\paragraph{Reachability.} We build the inter-procedural call graph once and take
the forward closure from the dispatch routines, recording for every recovered
function whether it is reachable from a dispatcher and at what distance. The
graph edges are persisted, not discarded after the closure.

\paragraph{Controlled-input taint.} A lightweight taint pass over
\glaurung's lifted intermediate representation~\cite{glaurung} seeds one
attacker-controlled root at the
dispatch routine, the buffered input pointer
\code{Irp->AssociatedIrp.SystemBuffer} (at \code{IRP}+\code{0x18} on x64), and
records each sink where that value reaches a dereference or store, tagged with
the base register, displacement, and whether a comparison sanitizes it on the
path. The pass is intra-procedural over a fixed window from the dispatch
routine, and every reported sink lies in the dispatch body. This source is the
buffered and direct-I/O input, roughly 40\% of the 64-bit-lane codes by transfer
method. The \code{METHOD\_NEITHER} and direct-I/O data-buffer paths are out of
scope (Section~\ref{sec:discussion}). The unsanitized sinks are the precise
signal the downstream stages use. A structural bug-class pre-pass over every
function supplies a coarser, higher-recall lead set.

\paragraph{Surface and provenance.} A permissive device security descriptor is
read \emph{from the binary} (an embedded SDDL granting write access to a
low-privilege SID, with no process-trust label) and flagged as a high-recall
prefilter for unprivileged reachability. PE version information (company,
product, version, original filename), the compile timestamp, the
Authenticode-signed flag, and the harvest-path vendor are recorded. Imports and
exports are captured as a feature vector. The prefilter is imperfect in both
directions: it misses drivers whose device ACL is set at runtime, by an INF
\code{AddReg}, or by a device-class default (a known false negative is our own
\code{NDKPing} test driver, reachable but with no SDDL in its image), and it can
over-flag when an embedded SDDL binds a different object than the control device.
Confirmed reachability requires the applied descriptor, not the embedded string.
Functions, call edges, imports, and exports are written for every driver under a
size budget (Section~\ref{sec:eval}).

The Recover pass is deterministic and reproducible: identical bytes and tool
versions yield identical rows. It is bounded, not exhaustive. The dispatch and
taint backends target the WDM and index-dispatch shapes. The framework-driver
(KMDF/WDF) callback path is recovered for structure, but its retrieve-buffer
taint source is future work, so WDF leaf handlers carry structure without
precise sinks today.

\subsection{Enrich}
\label{sec:enrich}
The model stage runs only on the prioritized tier: dispatch routines in drivers
behind a permissive device descriptor that carry at least one unsanitized
buffered-input sink. For each, we build a grounding context from the recovered facts, never
from decompiled bytes: the device access-control descriptor, the decoded IOCTL
codes with their transfer methods, the taint sinks with their offsets and
sanitization flags, the high-tier structural bug-class hits, and a
ground-truth disassembly excerpt of the dispatch head. We prompt an LLM (GPT-5.5 or
Claude Opus 4.7, via \code{pydantic-ai} for schema-enforced output) for a
structured verdict: a risk level, a
primitive (out-of-bounds read/write, type confusion, arbitrary read-write,
none), an attacker-reachability tier, a CWE set, a one-line role, and a
rationale that must cite the deciding sink. The model is instructed that the
attacker controls the input buffer and its lengths and that the privilege
follows the device descriptor.

Two design choices keep this faithful and cheap. The context is the taint slice
plus disassembly, never lifted or decompiled C, because enriched pseudo-code is
a ranking seam, not ground truth: it hallucinates truncations and drops guard
fields, and a verdict built on it inherits those errors. The model is also
gated by the deterministic tier, so the expensive component runs on the dozens
of high-value handlers, not the millions of functions. The structural bug-class
pre-pass that supplies leads runs on a cheaper model.

\subsection{Rank}
\label{sec:rank}
The store is queried through relational views. The hunt view orders
user-reachable dispatch routines that carry unsanitized controlled-input sinks
or high-tier bug-class hits and are not already covered by a filed finding,
ranked by unsanitized-sink count. A coverage view rolls up, per corpus build
and architecture, how much surface was recovered, how much is
unprivileged-reachable, and how much is enriched. A cross-driver view groups the
same decoded \code{(device\_type, function\_number)} across distinct drivers and
builds, surfacing shared or copied control interfaces. An oracle-agreement view
joins the static taint sinks against an ingested set of symbolic-execution
results at the same function, so the two methods can check each other. The
views read the store read-only and are safe to run while a build is still in
progress.

\begin{table}[t]
  \centering
  \small
  \begin{tabular}{@{}ll@{}}
    \toprule
    \multicolumn{2}{@{}l}{\textbf{Store schema}\quad key \code{(binary\_sha256, function\_va)}} \\
    \midrule
    \emph{Recover} & \code{binaries}, \code{dispatchers}, \code{ioctl\_codes}, \code{handlers}, \\
                   & \code{functions}, \code{call\_edges}, \code{imports}, \code{exports}, \\
                   & \code{taint\_sinks}\textsuperscript{\dag}, \code{bugclass\_hits}\textsuperscript{\dag} \\
    \emph{Enrich}  & \code{handler\_enrichment}\textsuperscript{\dag} \\
    \emph{Ingest}  & \code{ioctlance\_vulns}\textsuperscript{\dag}, \code{findings}\textsuperscript{\dag} \\
    \midrule
    \multicolumn{2}{@{}l}{\textbf{Views}} \\
    \code{v\_unaudited\_unpriv\_surface}\textsuperscript{\dag} & ranked hunt worklist \\
    \code{v\_coverage}                   & recovered/reachable/enriched per build \\
    \code{v\_cross\_build}                & shared \code{(device\_type, function)} across drivers \\
    \code{v\_oracle\_agreement}\textsuperscript{\dag}           & taint sinks vs.\ ingested symbolic results \\
    \bottomrule
  \end{tabular}
  \caption{The relational store. Every stage writes rows keyed by
  \code{(binary\_sha256, function\_va)}; the four views read it read-only.
  Persisting the structure, rather than emitting a finding list, is what enables
  the cross-driver and oracle-agreement queries. Rows marked
  \textsuperscript{\dag} are the targeting tier and are withheld from the public
  dataset~\cite{ioctlcensusdata}; the remaining structural rows are released.}
  \label{tab:schema}
\end{table}

\section{What the Pipeline Recovers}
\label{sec:eval}

We characterize the database, not a detection rate: how much surface the
deterministic pass recovers, how the model tier narrows it, and where the
pipeline stops. Table~\ref{tab:corpus} gives the corpus totals.

\begin{table}[t]
  \centering
  \small
  \setlength{\tabcolsep}{6pt}
  \begin{tabular}{@{}l r r@{}}
    \toprule
                                     & \textbf{Count} & \textbf{\% corp.} \\
    \midrule
    \multicolumn{3}{@{}l}{\textbf{Coverage} of 27{,}087 signed drivers} \\
    Dispatch surface recovered       & 21{,}708       & 80.1 \\
    \quad x64                        & 10{,}781       & 39.8 \\
    \quad x86, surface-only          & 10{,}927       & 40.3 \\
    No recovered surface, or capped  & 5{,}379        & 19.9 \\
    \addlinespace
    \multicolumn{3}{@{}l}{\textbf{Recovered surface} (x64 + x86)} \\
    Decoded control codes            & 3{,}089{,}633  &      \\
    Primary dispatch routines        & 21{,}708       &      \\
    Handlers                         & 848{,}094      &      \\
    \addlinespace
    \multicolumn{3}{@{}l}{\textbf{Per-function layer} (x64 lane only)} \\
    Functions                        & 8{,}184{,}655  &      \\
    Call edges                       & 15{,}950{,}899 &      \\
    Buffered-input taint sinks       & 63{,}263       &      \\
    \bottomrule
  \end{tabular}
  \caption{The corpus. A driver counts as \textbf{surface recovered} when a
  primary \code{IRP\_MJ\_DEVICE\_CONTROL} dispatch routine and its decoded codes
  are recovered; one architecture-neutral pass reaches $80\%$ across x64 and x86
  (Section~\ref{sec:multiarch}). The x64 lane is analyzed deeply ($12{,}719$
  binaries; $10{,}781$ yield a dispatch surface); the x86 lane is surface-only,
  so functions, call edges, and taint sinks are the x64 lane alone.
  The decoded-code total is heavy-tailed (Figure~\ref{fig:codedist}): the legacy
  x86 population carries large NDIS-style request maps.}
  \label{tab:corpus}
\end{table}

\paragraph{The deterministic pass is the artifact.} Across the \nx\ fully
analyzed 64-bit drivers the \emph{Recover} stage produced \nfns\ functions,
\nedges\ call edges, \ncodes\ decoded IOCTL codes, and \nsinks\ buffered-input
taint sinks. It recovers a primary dispatch routine, the
\code{IRP\_MJ\_DEVICE\_CONTROL} handler, for \nprimary\ of those drivers (0.85
per driver); a further \ndisp\ \emph{dispatch-shaped} functions are recorded,
the sub-dispatchers and single-constant sites that a one-best-per-driver tag
separates from the primary. These persist together, so the store answers
structural queries a finding list cannot: the cross-driver view groups a single
decoded \code{(device\_type, function\_number)} across distinct vendors, and the
reachability column distinguishes a handler one call from the dispatcher from
one buried ten edges deep. None of this requires the model.

\paragraph{One row, end to end.} A single store row threads the layers a finding
list collapses into one line. For the test driver \code{NDKPing} the store holds
its provenance, the device descriptor, the decoded control code \code{0x220404}
as \code{(device\_type 0x22, function 0x101, METHOD\_BUFFERED)}, and the precise
unsanitized sink, a controlled dereference at \code{SystemBuffer}+\code{0x28} with
no dominating check. The same row makes the prefilter limitation concrete: the
sink is recovered exactly, yet the runtime-applied descriptor keeps the driver
out of the reachable tier. On a world-writable production driver a structurally
similar entry, a permissive descriptor over a recovered sink, ranks near the top
of the order: the \code{nvlddmkm.sys} surface of Section~\ref{sec:backtest}.

\paragraph{Coverage is bounded by design, and the bounds are recorded.} The
pipeline does not analyze every byte of every driver. Two limits dominate, and
both are written into the \code{analysis\_level} column rather than hidden. First, the deep per-function work (the full disassembly that
feeds the call graph and bug-class scan) is gated by a size cap: the corpus
includes drivers up to 106\,MB, and a single 10\,MB driver peaks at roughly
13\,GB of resident memory under full disassembly, so drivers above 4\,MB keep
their IOCTL surface and taint but skip the full-function layer, and drivers
above 20\,MB are recorded with provenance only. This affects under five percent
of the corpus and every capped driver is a named, re-runnable batch. Second,
depth varies by architecture: the x64 lane is analyzed deeply (call graph and
taint), while the x86 lane that the architecture-neutral recovery added
(Section~\ref{sec:multiarch}) is surface-only, recovering dispatch routines and
decoded codes without the per-function layer. What remains provenance-only is
the smaller residue: framework and router drivers with no in-binary control-code
compare, the arm64 tail, and a handful of 16/32-bit relics.

\begin{figure}[t]
  \centering
  \includegraphics[width=\columnwidth]{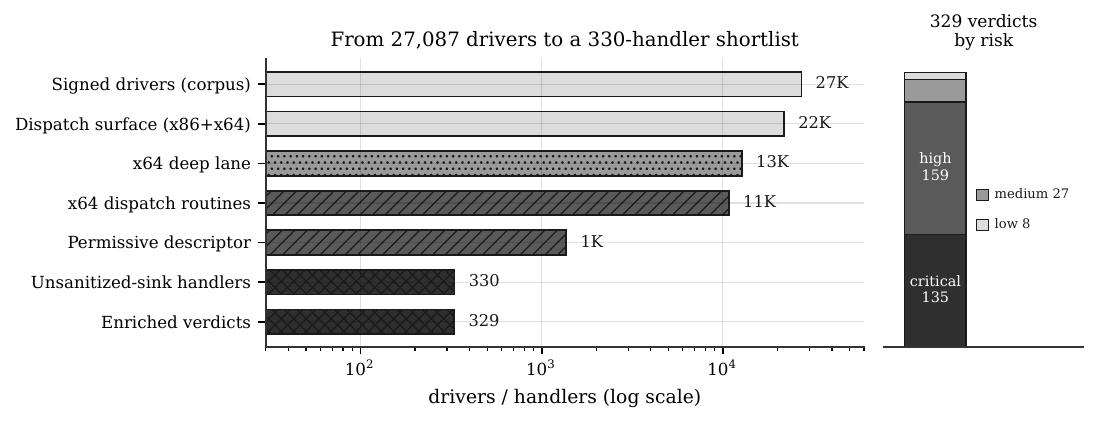}
  \caption{The deterministic filter, not the model, does the narrowing. From the
  27{,}087-driver corpus to a 330-handler shortlist: surface recovery, then within
  the deeply analyzed x64 lane the permissive-descriptor and unsanitized-sink
  tiers. Right: the \nenriched\ completed verdicts by risk. Log scale.}
  \label{fig:funnel}
\end{figure}

\paragraph{The model tier is narrow and selective.} \emph{Enrich} ran on the
precise tier: dispatch routines behind a permissive device descriptor carrying
an unsanitized buffered-input sink, $330$ handlers across $228$ drivers. Because
the unsanitized-sink filter needs taint, the model tier draws from the deeply
analyzed x64 lane specifically. The deterministic filter, not the model, does
the heavy narrowing: from the \nprimary\ x64 dispatch routines, to $1{,}366$
behind a permissive descriptor, to the $330$ handlers that
also carry an unsanitized sink. We enriched \nenriched\ of the $330$ (one was not
resolvable to a path). The model is liberal with its top label: it rates $135$ of
the $329$ critical and $159$ high (Figure~\ref{fig:funnel}, right), a $41\%$
critical rate that marks the tiers as prioritization hypotheses, not calibrated
severities.

\begin{table}[t]
  \centering
  \small
  \begin{tabular}{lr}
    \toprule
    \multicolumn{2}{l}{\textbf{(a) \code{ioctlance} native run, 1{,}000-driver sample}} \\
    \midrule
    Sampled (surface-recovered x64)      & 1{,}000 \\
    Completed within 40\,s               & 802 \\
    \quad timed out / crashed            & 198 \\
    Discovered an IOCTL handler          & 126 \;(15.7\% of completed) \\
    Reported $\geq$1 vulnerability state  & 54 \\
    \midrule
    \multicolumn{2}{l}{\textbf{(b) Cross-method agreement (802 completed)}} \\
    \midrule
    Our taint: drivers with unsanitized sink & 231 \\
    \code{ioctlance}: drivers with a vuln     & 54 \\
    \quad both methods flag the driver        & 35 \;(65\% of \code{ioctlance}'s) \\
    \quad \code{ioctlance} only               & 19 \\
    \quad taint only                          & 196 \\
    \qquad of which \code{ioctlance} reached no handler & 171 \\
    \bottomrule
  \end{tabular}
  \caption{Cross-method run. \code{ioctlance} is run in its native
  handler-discovery mode (an independent oracle, not fed our handler
  addresses). Panel (a) is its coverage on the population where our
  deterministic pass already recovered a dispatch surface; panel (b) joins its
  findings to our static taint over the drivers it finished. Agreement is
  driver-level co-flagging, not confirmed bugs: neither method is ground truth.}
  \label{tab:oracle}
\end{table}

\paragraph{Cross-method agreement, measured.} To test the claim that the store
lets two methods check each other, we ran \code{ioctlance}, a whole-driver
symbolic-execution scanner~\cite{ioctlance}, in its native discovery mode over a
random \nsamp-driver sample of the surface-recovered x64 set, and joined its
findings to our static taint (Table~\ref{tab:oracle}). We used our own
fork~\cite{ioctlancefork}, which refactors the original for parallel batch
execution and adds time and space bounds for the out-of-memory and
pathological-runtime cases whole-driver symbolic execution hits at this scale. The first result is coverage. On drivers
where our deterministic pass had already recovered a dispatch surface, the
symbolic engine's front-door handler discovery reached \nioccov\ of
\nioccompleted\ at a forty-second budget (a further $198$ timed out), all on our
own recovered population. A budget sweep shows this is near the engine's
native-discovery ceiling rather than a clock artifact: on a matched subsample,
five- and fifteen-minute budgets lift handler discovery by two drivers and then
none, more than halving the timeout fraction, while completed runs finish in
about a minute at every budget. The residual gap is x64-only path explosion. Where it did fire, the two methods agree: of the \niocvd\ drivers
\code{ioctlance} flagged, our taint independently flagged \nboth, a $65\%$
driver-level overlap. That overlap sits entirely on buffered and direct-method
codes: no \code{METHOD\_NEITHER} code appears in the agreement set, against an
eleven percent share of the codes \code{ioctlance} flags alone.

\paragraph{Disagreement is reach, not contradiction.} The static pass flags far
more drivers, \ntaintd\ to \niocvd, and the asymmetry is almost entirely the
symbolic engine's reach ceiling rather than a precision gap: of the $196$
drivers our taint flags alone, \code{ioctlance} discovered no handler for
\ntonlynoh\ of them. It cannot disagree with a sink in a handler it never
entered. The reverse disagreement is the more useful half. The $19$ drivers
\code{ioctlance} flags alone are enriched in exactly our two recorded blind
spots: \code{METHOD\_NEITHER} codes, which the buffered source does not seed,
and double-free states, a temporal-safety class a spatial taint-to-sink pass
does not model at all. Ingesting the symbolic result is therefore not redundant
with the taint; it backfills the surface the taint is blind to by construction.
Neither tool is ground truth, and the symbolic engine's double-free class is its
most false-positive-prone, so agreement raises confidence rather than confirming
a defect. Keeping both columns and exposing the join is the capability this
comparison shows.

\paragraph{The leads are leads.} The \nenriched\ verdicts include $135$ rows the
model rated critical, $129$ of them with a model-assigned arbitrary read-write
primitive, on user-reachable devices. These are candidate needles for a
verification pass, not confirmed bugs. We report no true-positive rate. Several of the
top-ranked drivers (wireless-LAN parts from common chipset vendors) overlap
targets our prior triage already retired as scooped or hardened, the expected
outcome: the database ranks the surface, and a separate disassembly-and-reverify
pass, with a freshness-checked public-disclosure search, decides each one.

\section{Recovery Across Architectures}
\label{sec:multiarch}

\begin{figure}[t]
  \centering
  \includegraphics[width=\columnwidth]{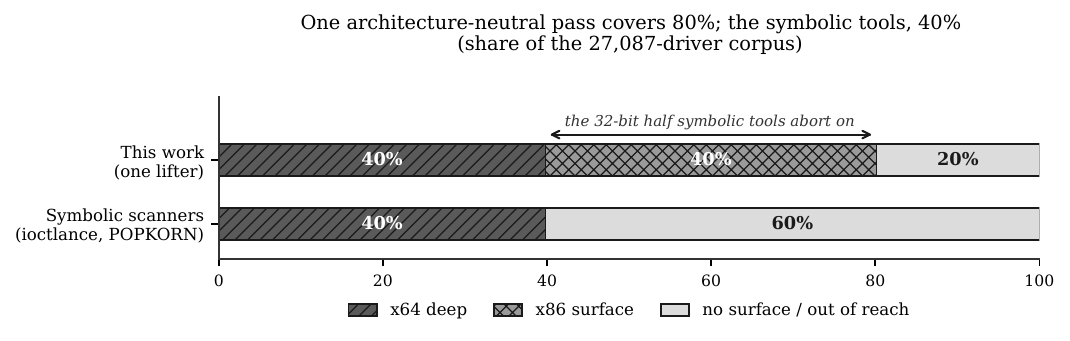}
  \caption{Control-code-surface coverage of the corpus. One architecture-neutral
  pass recovers a dispatch surface for $80\%$ of the 27{,}087 drivers across x64
  and x86; the symbolic scanners are x64-only and abort on the 32-bit half, which
  caps them near $40\%$. The empty x86 region of the lower bar is the
  contribution.}
  \label{fig:coverage}
\end{figure}

\paragraph{Surface recovery generalizes; the symbolic tools do not.} Roughly half
the corpus is 32-bit x86, much of it the legacy vendor drivers the
bring-your-own-vulnerable-driver class draws on. The symbolic scanners cannot
reach it: in our runs \code{ioctlance} aborts on every 32-bit driver with an
x86-64-only control-register error, and POPKORN and ScrewedDrivers target x64. Our dispatch
recovery has no such bound, because it reads \glaurung's architecture-neutral
intermediate representation. The cmp-immediate dispatch shape, a handler
comparing the control code against decoded constants, lifts to the same IR
statements regardless of instruction set, so one recognizer recovers the surface
on both x86 and x64 (arm64 lifts cleanly too, but defers code recovery for the
reason in the last paragraph). Enumerating function starts is the only
architecture-specific step: x64 and arm64 from the exception table, x86 from
relocation-derived code pointers, since 32-bit images carry no unwind data.

\begin{figure}[t]
  \centering
  \includegraphics[width=\columnwidth]{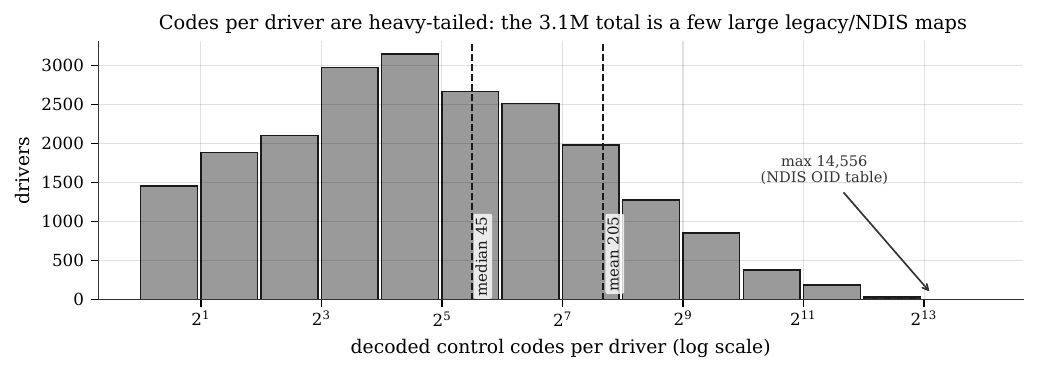}
  \caption{Decoded control codes per driver, over the surface-recovered drivers
  that carry at least one code. The distribution is heavy-tailed: median $45$,
  mean $205$, with a tail reaching $14{,}556$ in a single NDIS OID-table driver.
  The $3.1$M aggregate is dominated by a few large legacy and NDIS maps, which is
  why the headline coverage metric counts drivers with a recovered surface, not
  raw codes.}
  \label{fig:codedist}
\end{figure}

\paragraph{The x86 recovery is disassembly-confirmed.} On the 32-bit lane the
codes we recover are precise, not complete: across a
stratified sample every control code the IR recovery reports is independently
confirmed by a capstone disassembly of the same dispatcher ($272$ of $272$, a
precision check, not a recall bound), and the per-driver
transfer-method and device-type distributions match real IOCTL conventions
(buffered-dominant, custom device types) rather than the uniform spread of
incidental constants. We also reject a false-positive class the cross-check
surfaced: four-byte printable-ASCII signatures (driver ``magic'' tags) that pass
a naive code-shape test. Recovery reaches a dispatch surface on \nxsixrec\ of
the $13{,}105$ x86 drivers (\xsixrecall); the misses are framework drivers and
pure routers with no in-binary control-code compare. Through this lane \nxsixrec\ x86 drivers and \nxsixcodes\
decoded codes join the database, taking control-code-surface coverage from the
64-bit \covbefore\ to \covafter\ of the corpus. The legacy x86 population is
network-heavy: $17\%$ of the recovered drivers link NDIS, and for those the
recovered constants are object-identifier (OID) request codes rather than
\code{DeviceIoControl} IOCTLs. They are the same kind of artifact, a control code
the dispatcher compares against, and the same recognizer finds them. This is why
we describe the surface as control-code rather than IOCTL-only. It is also why the
code total is heavy-tailed (Figure~\ref{fig:codedist}): a reader after IOCTL
specifically should filter on the recorded driver model. We run the x86 lane
surface-only: it recovers the dispatch routines and decoded codes, and the
heavier per-function call-graph and taint layer stays on the 64-bit set, where
the precise signal is validated.

\paragraph{What does not yet generalize.} Surface recovery is
architecture-portable; the precise taint signal is not. x86 passes the IRP on the stack rather than in a register, and unlike
the dispatch codes its sinks have no independent oracle, since \code{ioctlance}
cannot run on 32-bit drivers to cross-check. x86 therefore contributes dispatch
routines and decoded codes but not validated taint sinks; the 32-bit taint source
is implemented and gated behind a confirmation that the candidate is a real
\code{DRIVER\_DISPATCH}, but we do not persist its sinks until they can be
calibrated against ground truth. arm64 lifts cleanly but materializes each 32-bit
control code through a \code{mov}/\code{movk} pair decoupled from the compare,
which the constant-folding recognizer does not yet follow, so arm64 stays
provenance-only. Both are recorded as open work.

\section{A Known-CVE Backtest}
\label{sec:backtest}

A ranking should place known-vulnerable handlers near its top. As a smoke test,
not a recall measurement, we ran a backtest against four oracles: the LOLDrivers
catalog~\cite{loldrivers},
public CVE writeups, the Microsoft vulnerable-driver blocklist, and our own
prior confirmed findings, where we know the vulnerable driver, the IOCTL code,
and the handler address exactly. For each oracle we matched its drivers to the
corpus by filename, then measured what the deterministic pass recovered and
where the driver fell in the hunt order. The result is one clean positive and a precise map of where the signal is blind.

\paragraph{The public corpora barely intersect ours.} Our corpus is dominated by
Microsoft Update Catalog drivers; the famous bring-your-own-vulnerable-driver
binaries are third-party installer payloads, and most are simply absent.
Table~\ref{tab:backtest} gives the overlap: of dozens of catalogued vulnerable
drivers per oracle, four to seven match by name, and fewer still recover a
surface. A meaningful public-CVE recall measurement would require ingesting an
actual vulnerable-driver corpus, not the Update Catalog.

\begin{table}[t]
  \centering
  \small
  \begin{tabular}{lrrr}
    \toprule
    \textbf{Oracle} & \textbf{In corpus} & \textbf{Surface} & \textbf{Rankable} \\
    \midrule
    LOLDrivers              & 4 & 0 & 0 \\
    Public CVE writeups     & 6 & 5 & 1 \\
    MS blocklist / BYOVD    & 7 & 4 & 0 \\
    Our confirmed findings  & 6 & 6 & 5 \\
    \bottomrule
  \end{tabular}
  \caption{Backtest overlap. \textbf{In corpus} is a filename match against the
  \nbins\ drivers; \textbf{Surface} also recovers a dispatch surface on top of
  that match; \textbf{Rankable} further passes the permissive-descriptor and
  unsanitized-sink gates and so appears in the hunt order. The public oracles
  match a handful each (the corpus is Update-Catalog drivers, not a
  bring-your-own-vulnerable-driver set); our own findings match by construction.}
  \label{tab:backtest}
\end{table}

\paragraph{Where a CVE is in scope, the ranking is sharp.} The one public CVE
that is present, buffered-method, and behind a world-writable device is
\code{nvlddmkm.sys}, whose display-driver escape surface carries public CVEs such
as CVE-2024-0090. Its highest-ranked reachable handler sits at 27th of $19{,}596$
rows, the top $0.14\%$. The \code{NDKPing} test driver's documented
\code{SystemBuffer}+\code{0x28} sink is recovered exactly. Within the slice the
taint source models, the order concentrates known-bad at the top.

\paragraph{The misses are structural, and we name each one.}
Against our six own confirmed true positives the taint pass recovered the
vulnerable sink for only one (\code{NDKPing}); the other five are
framework-model (WDF/KMDF) drivers whose input arrives through
\code{WdfRequest\allowbreak Retrieve\allowbreak InputBuffer}, not \code{SystemBuffer}, so the WDM source
never fires. The coarser bug-class proxy partly compensates: it lands a
high-tier hit on the exact vulnerable handler for three of those five. The
public-oracle misses have the same shape: most blocklist drivers expose a
physical-memory map (\code{MmMapIoSpace}) or model-specific-register primitive
rather than a buffer overflow, which a buffered-input taint source cannot
represent. One recent entry makes the limitation concrete: the driver behind
CVE-2025-8061 is in the corpus at a vulnerable build with its exact IOCTL code
decoded, yet the buffered-input signal never surfaces it. Its IOCTLs are a
register read-write, not a buffer flaw, so the taint source models no sink.

\paragraph{Two failures the backtest surfaces.} The unprivileged-reach
prefilter and the precise taint signal are disjoint on the one driver where both
should fire: \code{NDKPing}'s sink is recovered, but its device descriptor is set
at runtime, so the prefilter marks it unreachable and drops it from the tier. And
the model rates $135$ of the $329$ enriched handlers \code{critical}. We
re-adjudicated that full tier against public disclosure status: $78$ re-flag
drivers that already carry a public CVE, vendor advisory,
LOLDrivers~\cite{loldrivers} entry, or published IOCTL audit (Synaptics, NVIDIA,
FTDI, and Intel, Realtek, and Qualcomm/Atheros wireless parts); $12$ more are
Ralink/MediaTek handlers on a heavily-audited bring-your-own-vulnerable-driver
surface; and $6$ are the \code{mlx5} driver our own triage already recorded as a
false positive. Only $39$ handlers, across $35$ mostly obscure single-vendor
USB-filter, HID, and NIC drivers, surfaced no public disclosure, and none is
confirmed as a novel defect. The critical tier is thus overwhelmingly a
re-flagging of already-spent surfaces, not a source of fresh targets. Both
failures have concrete fixes: a framework-model taint source, a
runtime-descriptor reachability check, and a penalty on already-audited binaries.

\section{Discussion}
\label{sec:discussion}

\paragraph{Relation to prior work.} The analysis techniques here are not new;
the artifact is. Symbolic scanners (ScrewedDrivers~\cite{screweddrivers}, POPKORN~\cite{popkorn},
\code{ioctlance}~\cite{ioctlance}) recover much the same surface but emit a
finding list and discard the structure, and they pay symbolic execution's path
explosion: in our cross-method run (Section~\ref{sec:eval}) \code{ioctlance}'s
native discovery reached a handler on $15.7\%$ of the runs it completed ($126$ of
$802$, all on drivers where our pass had already recovered one), under a fixed
per-driver budget that a further fifth of the sample exhausted. A budget sweep
(Section~\ref{sec:eval}) shows more time barely helps, so this is near the
engine's native-discovery ceiling. The durable contrast is architectural: the
symbolic front end is x64-only. VMware's Threat Analysis Unit pushed the same
symbolic approach to corpus scale in a 2023 retrohunt,
scanning roughly eighteen thousand drivers and triaging them down to a few dozen
vulnerable binaries~\cite{vmwaretau}, but it too was x64-only and emitted a
finding list rather than a queryable surface. Curated lists (LOLDrivers~\cite{loldrivers}, the Microsoft blocklist)
record the already-known few hundred drivers, not the long tail. A deeper
academic line analyzes driver code directly rather than cataloguing binaries:
Static Driver Verifier checks Windows drivers against API-usage
rules~\cite{sdv}, DR.~CHECKER runs flow- and context-sensitive taint over Linux
drivers~\cite{drchecker}, and DIFUZE fuzzes the Linux ioctl interface from
automatically recovered argument structures~\cite{difuze}. These are source-level
or Linux analyses; each finds bugs in a driver, and none builds a persistent,
cross-corpus map of the surface. On the
practitioner side, Winbindex indexes Windows binaries by build~\cite{winbindex}
and IRPMon traces live IRPs on a running system~\cite{irpmon}, but neither
decodes and stores the dispatch structure. Recent
LLM-on-binary work (VulBinLLM~\cite{vulbinllm}, kernel-driver triage
studies~\cite{csit}) labels individual binaries, and target-selection work
(\emph{Needles at Scale}~\cite{needles}, SiftRank~\cite{siftrank}) ranks
functions for analysis. We sit between them: a persistent, queryable database of
the IOCTL surface across a whole corpus, with a deterministic core and a thin
model layer, on which any of those methods can run. The combination is the gap none of them fill.

\paragraph{The closest concurrent work.} The nearest neighbor in scale and
timing is Threat Unpacked's scalable driver analyzer~\cite{threatunpacked}, a
pipeline that processes twenty-eight thousand drivers and, in a
second stage, tracks how the driver ecosystem evolves through automated
patch-diffing. It is an orchestration layer: it
wraps \code{ioctlance} for the analysis and \code{sigcheck} for provenance, and
persists the bug-finder's reports and patch-diff verdicts as taggable objects in
a malware-analysis database. We persist the recovered surface itself,
computed by our own recovery rather than delegated to a symbolic engine: the
decoded codes, dispatch routines, handlers, call graph, and taint sinks, keyed
by \code{(binary\_sha256, function\_va)}.
Three differences follow. Because the recovery is ours and architecture-neutral,
it covers the 32-bit lane the wrapped symbolic front end aborts on. Because the
codes are decoded and stored rather than carried as opaque constants, the
cross-driver view groups a shared decoded \code{(device\_type, function\_number)}
across vendors, an interface-identity axis distinct from the binary
code-reuse fingerprinting that pipeline uses to find copied handlers. And because
the store is the surface and not one tool's output, that pipeline's own analyses,
its patch-diff included, could run on top of it. The two are orthogonal: theirs
is a temporal, evolution-tracking view built on a bug-finder; ours is a static,
structural census of the surface that bug-finders can query.

\paragraph{What the database is good for.} Three uses follow from persisting
structure. A human or agent gets a ranked, deduplicated worklist instead of a
re-scan. The cross-driver view turns a single decoded control interface into a
cohort: a query for the drivers sharing one decoded \code{(device\_type,
function)} returns, for the most-shared interface in the corpus, $2{,}823$
distinct signed drivers across $130$ vendors, a span no per-binary finding list
can express and one the x86 lane widened by surfacing the legacy tail. And the
oracle-agreement view lets a static taint pass and a symbolic engine check each
other on the same store: they agree on $65\%$ of the drivers the symbolic engine
flags, and where they diverge each one's misses are the other's coverage.

\paragraph{Limitations.} Recovery is validated for precision, not recall: the
disassembly cross-check (Section~\ref{sec:multiarch}) confirms the codes and
dispatchers we report are real, but we do not measure how many we miss, and the
index-computed fallback can yield a partial or mis-rooted surface still counted as
recovered. Reported coverage is surface presence, not completeness, and mixes a
deep 64-bit lane with a surface-only 32-bit one. The x86 lane is surface-only
(Section~\ref{sec:multiarch}): it recovers dispatch routines and decoded codes,
validated against independent disassembly, but its taint sinks are not yet
persisted (the 32-bit stack-IRP source is implemented but lacks an independent
oracle to calibrate against), and arm64 stays provenance-only until the
\code{mov}/\code{movk} constant materialization is folded. The taint source on
the 64-bit lane seeds only the buffered
input pointer, so the \code{METHOD\_NEITHER} user pointer and the direct-I/O MDL
buffer are out of scope and a \code{METHOD\_NEITHER}-only handler yields no
sinks; the framework-driver retrieve-buffer source is likewise unimplemented, so
WDF leaf handlers carry structure without precise sinks. The
unprivileged-reachability flag is a static read of an embedded device descriptor
and is a prefilter, not a runtime fact, with false negatives where the ACL is
applied at runtime or by an INF. The model verdicts cover the unsanitized-sink
tier only, are hypotheses, and carry no true-positive rate. The backtest in Section~\ref{sec:backtest} bounds the effectiveness
claim: within the buffered, world-reachable slice the ranking is sharp (the one
in-scope public CVE lands in the top $0.14\%$), but the framework-driver and
physical-memory primitives that dominate the public corpora sit outside what the
present taint source models, and the public oracles barely intersect an
Update-Catalog corpus. Symbol names are resolved for only a minority of
functions, because the bulk join against an external symbol database did not
complete within our time budget; full resolution is a recorded backfill.

\paragraph{Ethics and data release.} The corpus is signed drivers that are
themselves publicly retrievable from the Microsoft Update Catalog and live
Windows installs, and the work is for defensive and authorized
vulnerability-research use. The census splits into a structural tier and a
targeting tier, and we release them differently. The structural tier, the
recovered surface itself, decoded codes, dispatch routines, handlers, the call
graph, imports and exports, and provenance, everything derivable from the
already-public binary, we publish as a dataset~\cite{ioctlcensusdata}: it records
that a device is reachable and which codes it decodes, not where an unchecked
write lives. Like any surface map it lowers the cost of analysis, but it points
at no specific primitive. The targeting tier
we withhold, the buffered-input taint sinks, the model verdicts, the structural
bug-class leads, the symbolic-execution results, and our own findings linkage,
because that join does localize candidate unprivileged-reachable primitives
across thousands of shipping drivers and is dual-use. The figures here are built
from aggregate distribution tables with no driver- or function-level data. The
machinery is open as well: the binary-analysis engine the pipeline is built on,
\glaurung~\cite{glaurung}, which supplies the lifting, disassembly,
decompilation, and intermediate representation underneath every result, is open
source, so the deterministic stages are reproducible against it. We release the
method, the structural census, the aggregate statistics, and the engine, and
hold back the targeting tier. Confirmed defects are handled through coordinated
disclosure outside this paper.

\section{Conclusion}
\label{sec:conclusion}

The Windows driver control-code surface has good scanners and one good list, but
no shared map. We built one: a census over \nbins\ signed drivers that recovers a
dispatch surface for \nsurf\ of them (\covafter), across x86 and x64 from one
architecture-neutral lifter, reaching the 32-bit half no symbolic scanner does.
It records, per driver and without symbolic execution, the dispatch routines and
decoded control codes, and on the 64-bit lane the handlers and their
reachability, the buffered-input taint sinks, the call graph, and signing
provenance, with a thin LLM layer over the handlers behind a
permissive descriptor. The persistent structure, not a one-shot finding list, is what makes the census
useful. A researcher facing tens of thousands of drivers no longer has to pick a
starting point by hand: one query groups a shared control code across vendors,
ranks the unaudited handlers behind a permissive descriptor, or lets a static
taint pass and a symbolic engine cross-check each other, all over the whole
corpus at once. We do not claim to have found every bug, or to have proven the
handlers the model flags; what we offer is a map that turns a vast, largely
unexamined attack surface into something a person or an agent can navigate and
prioritize. The remaining work deepens the same store rather than replacing it:
x86 taint sinks, arm64 dispatch, a framework-driver taint source, a
physical-memory primitive detector, and full symbol resolution. The schema is not
Windows-specific either, and a companion Linux IOCTL
Census~\cite{ioctlcensuslinux} applies the same method to the Linux ioctl
surface, so the two can be queried side by side.

We release the method, the structural census as a public
dataset~\cite{ioctlcensusdata}, the aggregate statistics, and the open analysis
engine, holding back only the targeting tier. A shared, queryable map of the
control-code surface should let the community spend its scarce
reverse-engineering effort where it counts, so that the next vulnerable handler
is reached by a query rather than by luck.

\bibliographystyle{plain}
\bibliography{bibtex/references}

\end{document}